\documentclass[twocolumn,showpacs,preprintnumbers,amsmath,amssymb]{revtex4}
\usepackage{graphicx}
\usepackage{dcolumn}
\usepackage{bm}
\usepackage{color}

\begin{document}

\title{Surface-plasmon based dispersive detection and spectroscopy of ultracold atoms}
\author{Matthias Mildner}
\author{Claus Zimmermann}
\author{Sebastian Slama}

\affiliation{Physikalisches Institut and Center for Quantum Science, Eberhard Karls Universit\"{a}t T\"ubingen, Auf der Morgenstelle 14, D-72076 T\"ubingen, Germany}
\date{\today}

\begin{abstract}
The paper reports on the optical detection and spectroscopy of ultracold atoms near a gold surface. A probe light field is used to excite surface plasmon polaritons. The refractive index of the atomic gas shifts the plasmon resonance and changes the reflected light power. Thus, the sensitivity of the detection is plasmonically enhanced. Absorption of photons from the evanescent wave is avoided by detuning the laser from atomic resonance which makes the detection scheme potentially nondestructive. The spectrum of the signal is determined by a Fano resonance. We show that atoms can be detected nondestructively with single atom resolution for typical parameters in cold atom experiments. Thus, the method is suitable for quantum nondemolition measurements of matter wave amplitudes. Experimentally, we measure a technically-limited sensitivity of 30 atoms and extend the detection scheme to dispersively image the atom cloud near the surface.
\end{abstract}

\pacs{}

\maketitle

\section{Introduction}

A quantum nondemolition (QND) measurement determines one of two conjugate variables of a quantum object with arbitrary precision. It is repeatable, i.e. the result of any succeeding measurement is exactly the same as the first result. Thus, the precision of QND measurements can exceed the standard quantum limit \cite{Caves1980}. A well established scheme in optics for the QND measurement of light field amplitudes is based on the Kerr effect leading to a crossed-phase modulation with a second (probe) light field \cite{Roch1992} whose phase is measured. A similar scheme has been proposed between matter waves and optical evanescent waves \cite{Courtois1995}. The matter wave amplitude, i.e. the number of atoms that are reflected at the evanescent wave can be determined in a QND measurement by analyzing the phase of the reflected light field. A more detailed analysis of the proposal including realistic experimental parameters has shown that a high signal-to-noise ratio can only be achieved if the evanescent wave is enhanced by resonant structures \cite{Aspect1995}. Since then, evanescent waves have been used in a series of experiments for reflecting and trapping cold atoms at plane surfaces with applications in matter wave optics and measurements of surface potentials \cite{Kasevich90, Aminoff93, Esslinger93, Seifert94, Westbrook98, Rychtarik04, Bender09, Bender10, Stehle11, Fiutowski13, Stehle14, Kawalec18, Mildner18}. How many atoms are trapped or moving in such experiments at very short, typically submicron distance from the surface is an intriguing question. This was first analyzed in \cite{Cornelussen02} where the resonant absorption of the evanescent wave by atoms at the surface was measured in the reflected light field. Subsequently the absorption of resonant light in the evanescent wave was used for measuring atom-surface interactions \cite{Ivanov04}. We note in this context that compared to the plane surface case much stronger absorptive and dispersive light matter interactions can been obtained by positioning cold atoms close to nanoscale devices as nanofibers \cite{Vetsch10, Dawkins11}, whispering gallery mode micro-resonators \cite{Alton11}, and photonic crystal waveguides \cite{Thompson13}, and single atom detection has been demonstrated in those and similar systems. We concentrate in this work on plane surfaces that are easier to fabricate and enhance the evanescent wave by exciting surface plasmon polaritons on a thin extended gold layer \cite{Kretschmann68}. The interaction of the narrow atomic resonance with the broad plasmonic resonance leads to a Fano lineshape in the spectrum \cite{Giannini11}. Such Fano resonances have been observed in ground-breaking experiments in atomic vapor cells \cite{Stern14, Talker18}. We extend this kind of spectroscopy to ultracold atoms where the linewidth of the Fano resonance is not limited by the Doppler broadening, but only by the natural linewidth of the atomic state. Thus, much higher sensitivities can be reached, and we investigate if the detection scheme can be used for a QND measurement of the atom number for experimentally realistic parameters. The fact that we can probe the atoms dispersively, i.e. detuned such that absorption of light can be neglected, is a prerequisite for such a QND measurement. The paper is organized as follows. Section II deals with the theoretical description of the detection scheme. It introduces the principle of the detection, analyzes the achievable sensitivity, derives necessary conditions for the method to be nondestructive, and determines the maximum number of atoms that can be detected in a QND measurement. Section III presents experimental spectra of ultracold atoms probed by surface plasmon polaritons and determines the resolution of the detection by analyzing the signal-to-noise ratio. Furthermore, it demonstrates that the detection scheme can be used to dispersively image the atom cloud.

\section{Theoretical description}
\begin{figure}
\scalebox{1}{\includegraphics{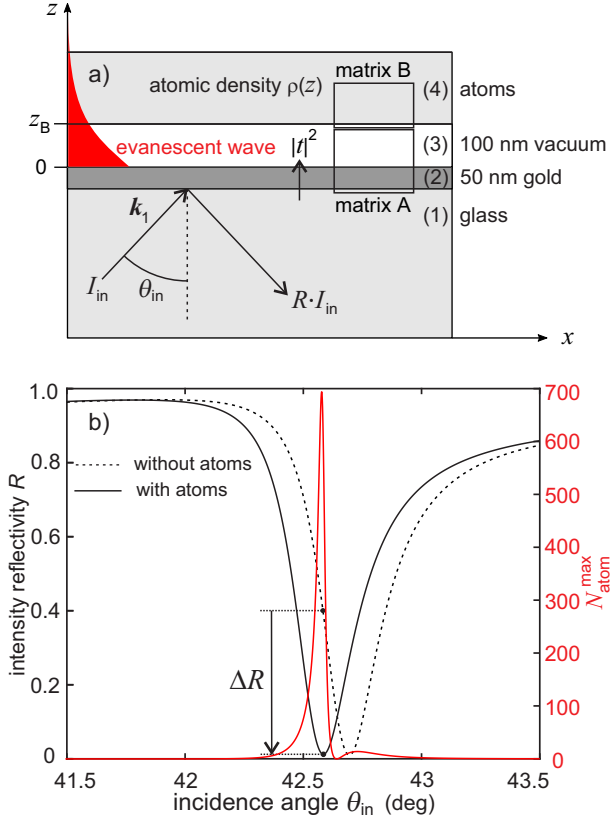}}
\caption{a) Experimental setup: A near resonant light field with intensity $I_\mathrm{in}$ and wave number $k_1$ is reflected by total internal reflection at a glass - gold - vacuum interface. It excites a plasmonically enhanced evanescent wave. Atoms are placed in a layer at a distance $z_B$ from the surface with atomic density $\rho\left(z\right)$. The interaction of the atoms with the evanescent wave changes the reflection of the light field. b) Intensity reflectivity $R$ as function of incidence angle $\theta_\mathrm{in}$ without atoms, i.e. $\rho = 0$, (black dashed curve), and for an atomic density of $\rho=1\cdot10^{13}~\mathrm{cm}^{-3}$ (black solid curve) at a detuning of $\delta=-30\,\Gamma$, and distance $z_B=100~\mathrm{nm}$. The maximum number of atoms that can be detected in a QND measurement is plotted in red. }
\label{fig:setup}
\end{figure} 
The experimental situation is described by the model shown in Fig.~\ref{fig:setup}a). A light field (probe laser) is reflected in Kretschmann configuration \cite{Kretschmann68} at a 3-layer-system consisting of glass (1), gold (2), and vacuum (3), and excites surface plasmon polaritons at the gold surface. This leads to a plasmonically enhanced evanescent wave that overlaps with an additional layer of ultracold Rubidium atoms (4) with atomic density $\rho\left(z\right)$ at a distance $z_\mathrm{B}$ from the gold surface. The gap between surface and layer (4) is caused by the action of the Casimir-Polder (CP) potential attracting the atoms towards the surface. For small distances the atoms are accelerated to the surface such that the atomic density is strongly reduced \cite{Stehle14}. For larger distances the CP potential can be compensated by additional external potentials as magnetic traps or repulsive evanescent wave dipole potentials \cite{Stehle11, Mildner18}. Thus, a potential barrier is generated that defines the minimum distance between atoms and surface \cite{Bender09}. The reflection of the light field from the layers is calculated with the transfer-matrix-method \cite{Deutsch94}. The 4-layer-system results in a total matrix $M_\mathrm{tot}=M_\mathrm{12}M_\mathrm{2}M_\mathrm{23}M_\mathrm{3}M_\mathrm{34}M_\mathrm{4}$, where $M_\mathrm{ij}$ describes the transition from layer $i$ to $j$ and $M_\mathrm{j}$ the free propagation inside layer $j$. In order to separate the effect of the atoms from properties of the solid layers, the total matrix is written as product
\begin{equation}\label{eq:Mtot}
\begin{split}
M_\mathrm{tot}&=A\cdot B\\\
&= \begin{pmatrix}
A_\mathrm{11} & A_\mathrm{12}\\
A_\mathrm{21} & A_\mathrm{22}
\end{pmatrix} \cdot \frac{1}{t_{34}}\begin{pmatrix}
e^{ik_\mathrm{4z}d_\mathrm{4}} & r_\mathrm{34}e^{-ik_\mathrm{4z}d_\mathrm{4}} \\
 r_\mathrm{34}e^{ik_\mathrm{4z}d_\mathrm{4}} & e^{-ik_\mathrm{4z}d_\mathrm{4}}
\end{pmatrix},
\end{split}
\end{equation}
with $A=M_\mathrm{12}M_\mathrm{2}M_\mathrm{23}M_\mathrm{3}$ and $B=M_\mathrm{34}M_\mathrm{4}$. The effect of the atom layer with thickness $d_4$ is included in the wave number z-component $k_{4z}$, and in the amplitude reflectivity $r_\mathrm{34}$ and transmission $t_{34}$ between layers (3) and (4). The imaginary $k_z$ number of the evanescent wave in layers $j=3,4$ is given by
\begin{equation}\label{eq:kz}
k_\mathrm{jz}=i k_\mathrm{0}\sqrt{n_1^2\sin^2\left(\theta_\mathrm{in}\right)-n_j^2},
\end{equation}
with free space wave number $k_0$, refractive index $n_j$ and incidence angle $\theta_\mathrm{in}$. The complex refractive index $n_4$ of the Rubidium gas is given by the Lorentz model and depends on the detuning $\delta$ of the laser frequency from the atomic transition. Reflectivity and transmission coefficients are derived from the Fresnel equations as
\begin{equation}\label{eq:rij}
r_\mathrm{ij}=\frac{k_\mathrm{iz}n_\mathrm{j}^2-k_\mathrm{jz}n_\mathrm{i}^2}{k_\mathrm{iz}n_j^2+k_\mathrm{jz}n_\mathrm{i}^2}
\end{equation}
and 
\begin{equation}\label{eq:tij}
t_{ij}=\frac{3k_\mathrm{iz}n_in_j}{k_\mathrm{iz}n_j^2+k_\mathrm{jz}n_\mathrm{i}^2}.
\end{equation}
The intensity reflectivity \cite{Deutsch94} of the layers is given by 
\begin{equation}\label{equa:refl}
\begin{split}
R&=|r_\mathrm{tot}|^2=\left|\frac{M_{tot}^{1,2}}{M_{tot}^{2,2}}\right|^2\\\
&=\frac{|A_\mathrm{11}|^2 |r_\mathrm{34}|^2+|A_\mathrm{12}|^2+2 \mathrm{Re}\left(A_\mathrm{11}{A_\mathrm{12}^*}r_\mathrm{34}\right)}{|A_\mathrm{21}|^2 |r_\mathrm{34}|^2+|A_\mathrm{22}|^2+2 \mathrm{Re}\left(A_\mathrm{21}{A_\mathrm{22}^*}r_\mathrm{34}\right)}
\end{split}
\end{equation}
and depends on the incidence angle in a characteristic resonant way, see Fig.~\ref{fig:setup}b). The minimum is caused by the excitation of surface plasmon polaritons. This resonance curve is shifted by the interaction with atoms in layer (4), depending on their density. For fixed incidence angle, atoms that are brought to the surface change the reflectivity by $\Delta R$. A measurement of the reflected light intensity can thus be used for detecting atoms close to the surface. The following paragraph deals with the sensitivity of this method.

\subsection{Sensitivity}
It is the refractive index $n_4$ of the atoms that leads to the change in reflection $\Delta R$. In general, $n_4$ is a complex number, as used in the simulation shown in Fig.~\ref{fig:setup}b). In the limit of large detuning $2\delta\gg \Gamma$ with natural linewidth $\Gamma$, it is mainly the real part of $n_4$ that is responsible for the occurence of $\Delta R$. This simplifies the following analytical description. Within the Lorentz model the real part of the refractive index can be expressed as
\begin{equation}\label{equa:linear}
n_4=1+\beta\cdot\rho,
\end{equation}
with atomic density $\rho$ and 
\begin{equation}\label{eq:beta}
\beta=\frac{3\lambda^3}{8\pi^2}\frac{\frac{2\delta}{\Gamma}}{1+\left(\frac{2\delta}{\Gamma}\right)^2}.
\end{equation}
The change in reflectivity for small atomic density is calculated from (\ref{equa:refl}) in first order Taylor expansion 
\begin{equation}\label{eq:DeltaR_1}
\Delta R =\frac{\partial R}{\partial \rho}\bigg\vert_{\rho=0}\cdot \rho
\end{equation}
by the partial derivative
\begin{equation}\label{eq:dR_drho}
\frac{\partial R}{\partial \rho}\bigg\vert_{\rho=0}=2\beta\left(1+\frac{k_\mathrm{0}^2}{2\kappa_3^2}\right)\frac{1}{|A_\mathrm{22}|^2}\Theta~,
\end{equation}
where the inverse decay length $\kappa_3=k_\mathrm{3z}/i$ of the evanescent wave in the gap was introduced. In the approximations made of small atomic density and large detuning, the refractive index $n_4\approx 1$. Thus, the inverse decay lenghts in the gap and in the atom layer are almost equal, i.e. $\kappa_3\approx\kappa_4$. The numerical factor 
\begin{equation}\label{eq:Theta}
\Theta=\left[\mathrm{Re}\left(A_\mathrm{11}A_\mathrm{12}^*\right)-\frac{|A_\mathrm{12}|^2}{|A_\mathrm{22}|^2}\mathrm{Re}\left(A_\mathrm{21}A_\mathrm{22}^*\right)\right]
\end{equation}
is typically a number on the order of one. An interesting aspect is revealed by substituting in Eq.~(\ref{eq:dR_drho}) $1/\vert A_{22}\vert^2=\vert t\vert^2\cdot\exp\left(-2\kappa_3 z_\mathrm{B}\right)$, where $\vert t \vert^2$ represents the plasmonic enhancement factor at the gold surface which is reduced by the exponential decay of the evanescent wave within the gap between surface and atoms. Thus, a large signal is favored by a large plasmonic enhancement and a small gap size. The change in reflectivity can now be expressed as
\begin{equation}\label{eq:DeltaR_2}
\begin{split}
\Delta R=&\frac{3\Theta}{4\pi^2}|t|^2\left(1+\frac{k_\mathrm{0}^2}{2\kappa_3^2}\right)
\frac{\frac{2\delta}{\Gamma}}{1+\left(\frac{2\delta}{\Gamma}\right)^2}\lambda^3\rho\exp\left(-2\kappa_3 z_\mathrm{B}\right).
\end{split}
\end{equation}
Equation (\ref{eq:DeltaR_2}) is valid for constant density $\rho$. It can be extended to include also an inhomogeneous density by integrating over $\rho(z)$ 
\begin{equation}\label{eq:DeltaR_3}
\begin{split}
\Delta R=&\frac{3\Theta}{2\pi^2}|t|^2\left(1+\frac{k_\mathrm{0}^2}{2\kappa_3^2}\right)	
\frac{\frac{2\delta}{\Gamma}}{1+\left(\frac{2\delta}{\Gamma}\right)^2}\lambda^3\kappa_3\cdot\\
&\cdot\int_{0}^{\infty}\rho\left(z\right)\exp\left(-2\kappa_3 z\right)\mathrm{dz}.
\end{split}
\end{equation}
It is apparent from (\ref{eq:DeltaR_3}) that the contribution of atoms at distance $z$ is weighted with the exponential decay of the evanescent wave. Thus, it is only a thin layer of atoms with thickness on the order of $l=1/\kappa_3$, i.e. typically a few hundred nanometers, that effectively contributes to the signal.\\ 

The signal $S$ is measured as change in the reflected light power, respectively in the number of reflected photons,
\begin{equation}\label{eq:signal}
S=\eta \Delta R N^\mathrm{}_\mathrm{in},
\end{equation} 
with detection efficiency $\eta$ and number of incident photons $N^\mathrm{}_\mathrm{in}$. The accuracy with which this signal can be detected is limited within the standard quantum limit by the Poissonian shot noise of detected photons
\begin{equation}\label{eq:shotnoise}
\Delta N_\mathrm{sql}=\sqrt{\eta R N^\mathrm{}_\mathrm{in}}.
\end{equation} 
The signal-to-noise ratio is thus given by
\begin{equation}\label{eq:SN1}
S_\mathrm{N}=\frac{\eta\Delta R N^\mathrm{}_\mathrm{in}}{\sqrt{\eta R N^\mathrm{}_\mathrm{in}}}=\sqrt{\frac{\eta N^\mathrm{}_\mathrm{in}}{R}}\Delta R.
\end{equation}
It is proportional to the change in reflectivity $\Delta R$ and increases with the number of incident photons. Thus, high photon numbers are in principle favorable for reaching a high signal-to-noise ratio. However, as described in the paragraph on nondestructiveness, the number of incident photons cannot be chosen arbitrarily large, as heating and loss of atoms increases with the light intensity. The maximum number $N^\mathrm{max}_\mathrm{in}$ of incoming photons under the demand of a maximum number $N^\mathrm{max}_\mathrm{abs}$ of absorbed photons is limited by (\ref{eq:N_ph_abs_2}). Together with (\ref{eq:DeltaR_2}) the signal-to-noise ratio is then given by
\begin{equation}\label{eq:SN2}
S_\mathrm{N}=\sqrt{\chi \lambda^3\rho \frac{\left(\frac{2\delta}{\Gamma}\right)^2}{1+\left(\frac{2\delta}{\Gamma}\right)^2} N^\mathrm{max}_\mathrm{abs}},
\end{equation}
with 
\begin{equation}\label{eq:chi}
\begin{split}
\chi=&\frac{3\eta}{2\pi^2} \Theta^2\frac{\sqrt{n_1^2\sin(\theta_\mathrm{in})^2-1}}{R} \left(1+\frac{k_\mathrm{0}^2}{2\kappa_3^2}\right)^2\\
&\cdot |t|^2 \exp\left(-2\kappa_3 z_\mathrm{B}\right).
\end{split}
\end{equation}
The signal-to-noise ratio in (\ref{eq:SN2}) is zero for $\delta=0$ and tends to a constant value for $2\delta\gg\Gamma$. This can be understood in the following way: for $\delta=0$, where the refractive index of the atoms equals the free space value, the signal is zero. For $2\delta\gg\Gamma$ the signal decreases for increasing detuning. But also the photon scattering rate decreases for increasing detuning, such that under the premise of a constant number of scattered photons, the useable light power increases with the detuning. The resulting resolution improves in a way that exactly cancels the dependence of $S_\mathrm{N}$ from the detuning. In conclusion, a detuning $2\delta\gg\Gamma$ is favorable for reaching a high signal-to-noise ratio.\\

\subsection{Nondestructiveness}
There are two major reasons why atoms are lost during the detection. One reason is the loss of atoms to the surface due to the attraction by the CP-potential. We assume that this loss mechanism is avoided by the generation of a potential barrier between the atom trap and the surface, for instance by far-detuned evanescent waves as demonstrated in \cite{Stehle11, Mildner18}. The second reason for atom loss is the absorption (and re-emission) of photons of near-resonant light, by which momentum is transferred to the atoms. In a quantum gas the momentum of a single photon is sufficient to depopulate the corresponding matter-wave. In a thermal cloud light absorption leads to heating that eventually lets atoms escape from the trap. It is thus important to determine the amount of absorbed light during the detection process. The energy flow of the evanescent wave is defined by the Poynting vector  
\begin{equation}\label{eq:Poyntingvector1}
\langle \vec S\rangle_\mathrm{t}=\mathrm{Re}\left(\frac{1}{2} \vec E \times \vec H^*\right),
\end{equation}
with transverse magnetic field
\begin{equation}\label{eq:HField}
\vec H = H_\mathrm{0}\hat{e}_\mathrm{y} \exp (-\kappa z)\exp (i(k_\mathrm{x}x-\omega t)).
\end{equation}
The Poynting vector of evanescent waves in nonabsorbing media is parallel to the surface. The absorption in the atomic gas is included in the imaginary part of the refractive index $n_4=n_\mathrm{4r}+i\cdot n_\mathrm{4i}$, which leads to an imaginary part of the inverse decay length $\kappa_4=\kappa_\mathrm{r}+i\cdot \kappa_\mathrm{i}$. Thus, the Poynting vector 
\begin{equation}\label{eq:S_t}
 \langle \vec S\rangle_\mathrm{t}=\frac{I_0}{k_\mathrm{0}}\cdot\exp\left(-2\kappa_\mathrm{r} z\right)
\begin{pmatrix}
k_\mathrm{x}\\
0 \\
2 n_\mathrm{4i} \kappa_\mathrm{r}-\kappa_\mathrm{i}
\end{pmatrix}
\end{equation}
in the limit of $n_\mathrm{4r}\approx 1$ and $n_\mathrm{4i}\ll 1$ has a vertical component, with evanescent wave intensity $I_0$ at the surface. The absorbed intensity per length is calculated by the divergence of the energy flow:
\begin{equation}\label{eq:dI_dr}
\begin{split}
\frac{dI_\mathrm{abs}}{dr}=&-\nabla \cdot \langle \vec S\rangle_\mathrm{t}\\
=&I_0 \frac{\left(4\kappa_r^2n_\mathrm{4i}-2\kappa_r\kappa_i\right)}{k_0}\exp(-2\kappa_rz).
\end{split}
\end{equation}
With $ \langle \vec S\rangle_\mathrm{t}$ given by (\ref{eq:S_t}), only the derivative in the $z$-direction leads to absorption.
Thus, the absorbed overall intensity is given by the integral of (\ref{eq:dI_dr}) over the $z$-direction 
\begin{equation}\label{eq:I_abs}
\begin{split}
I_\mathrm{abs}=&\int_{z_B}^{\infty}\frac{dI_\textrm{abs}}{dr}dz\\
=&I_0\frac{2\kappa_rn_\mathrm{4i}-\kappa_i}{k_0}\exp(-2\kappa_rz_B),
\end{split}
\end{equation}
where we have assumed a constant atomic density in layer (4). Relating the intensity $I_0=\vert t \vert^2 I_\mathrm{in}$ via the plasmonic intensity enhancement $\vert t \vert^2$ with the incoming intensity $I_\mathrm{in}$ , and expressing intensities by photon numbers, the fraction $f_N$ of absorbed to incoming photons is given by
\begin{equation}\label{eq:N_ph_abs}
f_N=\frac{N^\mathrm{}_\mathrm{abs}}{N^\mathrm{}_\mathrm{in}}=\frac{I_\mathrm{abs}}{I_\mathrm{in}}
=\vert t \vert^2\frac{2\kappa_rn_\mathrm{4i}-\kappa_i}{k_0}\exp(-2\kappa_rz_B).
\end{equation}
In order to determine the functional dependence of (\ref{eq:N_ph_abs}) on the detuning $\delta$ in a simple way, we make the approximation of $2\kappa_rn_\mathrm{4i}\ll\kappa_i$, and express $\kappa_i$ as a function of the refractive indices:
\begin{equation}\label{eq:kappa_i}
\kappa_i\approx -k_0\frac{n_\mathrm{4i}}{\sqrt{n_1^2\sin(\theta_\mathrm{in})^2-1}},
\end{equation}
with imaginary part of the refractive index of the atomic medium
\begin{equation}\label{eq:kappa_i}
n_\mathrm{4i}=\frac{3\lambda^3}{8\pi^2} \frac{\rho}{1+\left(\frac{2\delta}{\Gamma}\right)^2}.
\end{equation}
Thus, the fraction of absorbed photons is given by
\begin{equation}\label{eq:N_ph_abs_2}
\frac{N^\mathrm{}_\mathrm{abs}}{N^\mathrm{}_\mathrm{in}}=\frac{\vert t \vert^2}{\sqrt{n_1^2\sin(\theta_\mathrm{in})^2-1}}\frac{3\lambda^3}{8\pi^2}\frac{\rho\exp(-2\kappa_rz_B)}{1+\left(\frac{2\delta}{\Gamma}\right)^2}.
\end{equation}
This result lets us determine the maximum number of incident photons $N^\mathrm{max}_\mathrm{in}$ under the premise that a maximum number of $N^\mathrm{max}_\mathrm{abs}$ photons is absorbed, which is important for optimizing the signal-to-noise ratio in Eq. (\ref{eq:SN2}).\\

\subsection{Quantum nondemolition measurements}
\begin{figure}
	\scalebox{1}{\includegraphics{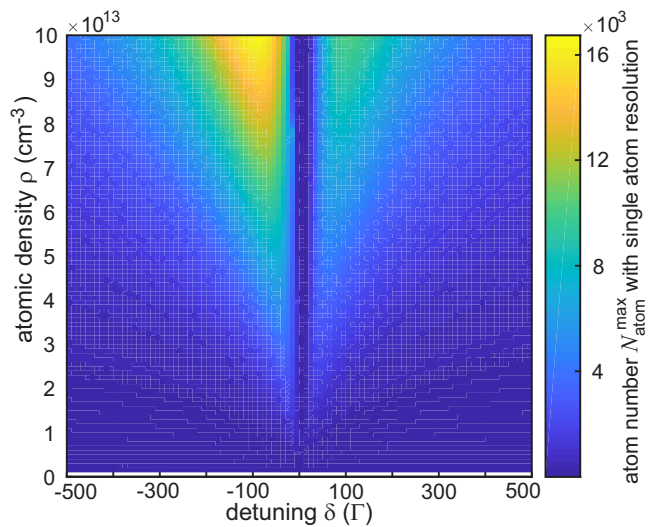}}
	\caption{The maximum number of atoms $N_\mathrm{atom}^\mathrm{max}$ that can be detected in a QND measurement increases with the atomic density $\rho$. For fixed density a maximum is reached for a certain detuning $\delta$.}
	\label{fig:QND}
\end{figure}

As proposed in \cite{Courtois1995, Aspect1995}, evanescent waves can in principle be used for the quantum nondemolition (QND) measurement of the amplitude of a matter wave that is reflected at the surface, i.e. the nondestructive measurement of the atom number with single atom precision. This paragraph shows that using the here presented method based on plasmonic enhancement the QND regime can be reached in typical cold atom experiments. We consider $N_\mathrm{atom}$ atoms that are confined in a volume $V=A\cdot l$ reaching from distance $z=z_B$ to $z=z_B+l$ vertically to the surface. The thickness $l=1/\kappa_4$ of the atom layer is limited by the range where the atoms contribute to the measured signal. The area $A$ parallel to the surface is given by the transverse size of the atom cloud. We assume that the beamwaist of the probe light field is matched to this area. The resolution in atom number $\Delta N_\mathrm{atom}$ is given by 
\begin{equation}\label{eq:resolution}
\Delta N_\mathrm{atom}=\frac{N_\mathrm{atom}}{S_\mathrm{N}},
\end{equation}
with signal-to-noise ratio $S_\mathrm{N}$. Single atom resolution requires $\Delta N_\mathrm{atom}\leq 1$. Thus, the maximum number of atoms that can be detected with single atom resolution is limited by
\begin{equation}\label{eq:Nmax1}
N_\mathrm{atom}^\mathrm{max}=S_\mathrm{N}=\sqrt{\frac{\eta N^\mathrm{}_\mathrm{in}}{R}}\Delta R,
\end{equation}
where we have inserted $S_\mathrm{N}$ from (\ref{eq:SN1}). The number of incident photons $N^\mathrm{}_\mathrm{in}$ is replaced by the number of absorbed photons $N^\mathrm{}_\mathrm{abs}$ using the fraction $f_N$ from (\ref{eq:N_ph_abs}). For the detection to be nondestructive, we demand that the number of absorbed photons should not exceed the number of atoms, i.e. we set
\begin{equation}\label{eq:Nmax2}
N^\mathrm{}_\mathrm{abs}= N_\mathrm{atom}^\mathrm{max}.
\end{equation}
Under this condition, $N_\mathrm{atom}^\mathrm{max}$ describes the maximum atom number where a QND measurement is possible:
\begin{equation}\label{eq:Nmax3}
N_\mathrm{atom}^\mathrm{max}=\frac{\eta}{f_N}\frac{\Delta R^2}{R}.
\end{equation}	
In order to calculate $N_\mathrm{atom}^\mathrm{max}$ correctly also for values of $\Delta R \sim 1$ where the reflectivity change gets nonlinear, Eq. (\ref{eq:Nmax3}) is not further analyzed analytically, but simulated numerically. 
The parameters in (\ref{eq:Nmax3}) depend on the incidence angle. We choose an atomic density of $\rho=10^{13}~\mathrm{cm}^{-3}$ typical in ultracold atom experiments, the wavelength $\lambda=780~\mathrm{nm}$ of the Rubidium $D2$ line, a detuning of $\delta=-30\cdot\Gamma$, detection efficiency $\eta=1$, and gap thickness $d_3=100~\mathrm{nm}$. The result for $N_\mathrm{atom}^\mathrm{max}$ as function of the incidence angle $\theta_\mathrm{in}$ is plotted in Fig.~\ref{fig:setup}b). The curve has a sharp maximum, which results from a combination of a maximum in the change in reflectivity $\Delta R$ with the fact that due to the action of the atoms the reflectivity at this specific angle is reduced to a very small value. Thus, the photon shot noise is strongly suppressed which improves the signal-to-noise ratio substantially. With the given parameters up to $N_\mathrm{atom}^\mathrm{max}=700$ atoms can be detected in a QND measurement. We have further analyzed how this number depends on the atomic density and on the detuning. The results are shown in Fig.~\ref{fig:QND}. We conclude that QND measurements are possible for typical atom densities $\rho$ used in cold atom experiments. The plot in Fig.~\ref{fig:QND} is not symmetric with respect to $\delta$, as the reflection curve $R(\theta_\mathrm{in})$ itself is asymmetric and shifts in opposite directions for positive and negative detuning, respectively. 

\section{Experimental detection of atoms}
The plasmon-based detection scheme is experimentally implemented by an ultracold atom experiment \cite{Stehle11, Stehle14} where a cloud of cold Rubidium atoms is prepared in a magnetic trap and can be moved and positioned with high precision using external magnetic fields. A detailed description of the setup can be found in \cite{Mildner18}. The atom cloud is initially prepared at a distance of $300~\mathrm{\mu m}$ from the surface of a glass prism on which thin square gold pads have been fabricated by atomic vapor deposition. The thickness of the gold pads of $d_2=40~\mathrm{nm}$ is measured by scanning force microscopy, and the sides of the squares are $200~\mu m$ in length.  After preparation and cooling of the atom cloud by forced radio-frequency evaporation to a temperature of $T\sim 1~\mathrm{\mu K}$ the atoms are moved towards the surface where they interact with the evanescent wave. The shape of the atom cloud is ellipsoidal with long $1/e^2$-radius $r_\mathrm{y}\approx 155~\mathrm{\mu m}$ and short radii $r_\mathrm{x}=r_\mathrm{z}\approx 50~\mathrm{\mu m}$). For detection of the atoms, a laser beam of an external cavity diode laser (ECDL) (the probe laser) is reflected by total internal reflection in the glass prism. The frequency of the probe laser field is controlled by beat-locking it to a reference ECDL which itself is stabilized using Rubidium saturation spectroscopy in a vapor cell. Thus, the detuning $\delta=\omega_\mathrm{L}-\omega_\mathrm{2,3}$ of the probe laser is adjustable in the range between $\pm20\cdot\Gamma$ around the $5s_{1/2},F=2 \rightarrow 5p_{3/2},F'=3$ transition. The probe laser is focussed onto the gold surface with beam waists of $\omega_\mathrm{x}=146\pm6~\mathrm{\mu m}$ and $\omega_\mathrm{y}=111\pm6~\mathrm{\mu m}$. The fact that the beam waist is larger in the $x$-direction is caused by the projection of the circular laser beam onto the surface under the angle of incidence. The position and the angle of incidence are adjusted by laser mirrors with an accuracy of $\Delta x=\Delta y=10~\mathrm{\mu m}$ and $\Delta\theta=~0.016^\circ$. After preparation of the atom cloud the magnetic trap minimum is moved in $5~\mathrm{ms}$ to a position which lies within the glass prism at $100~\mathrm{\mu m}$ distance from the surface. The atom cloud follows the trap minimum and reaches a velocity of $v_z=12.7~\mathrm{cm/s}$. The lateral position of the cloud is assured to coincide with the center of the evanescent wave. The probe laser is switched on only a short time ($\approx 4~\mathrm{ms}$) before the atom cloud reaches the surface in order to avoid heating by stray light at earlier times during the preparation of the atom cloud.
\begin{figure}[h]
	\centerline{\scalebox{1}{\includegraphics{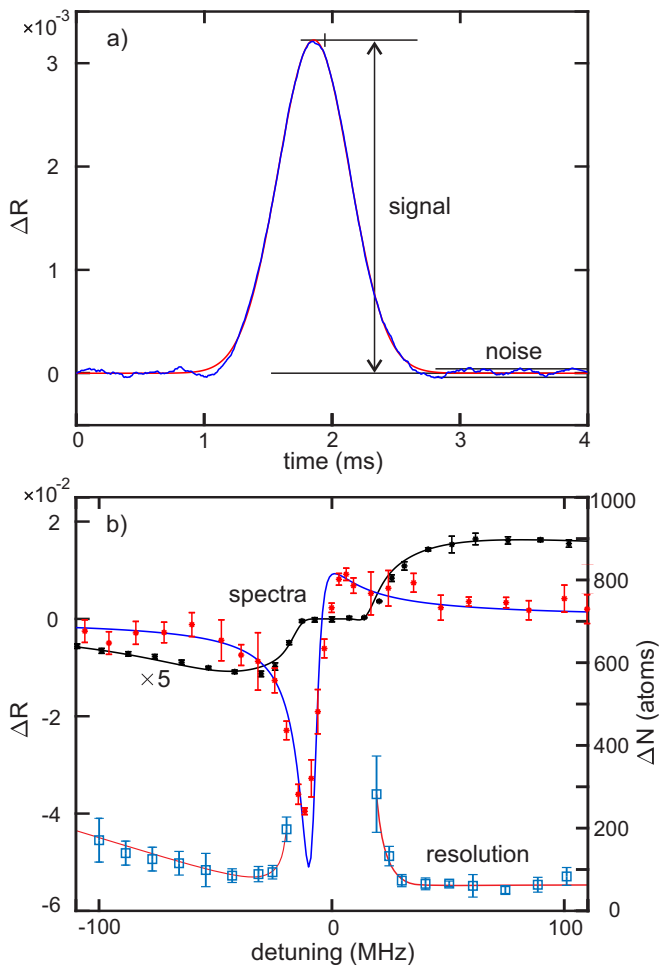}}}\caption{a) Measured change in reflectivity with detuning $\delta=12.6\cdot\Gamma$ and incident intensity $I_\mathrm{in}=14.2\cdot I_\mathrm{sat}$. The signal is fitted by a Gaussian function representing the density distribution of the cloud during its reflection at the surface. b) Measured and simulated spectra (asterisks). The black spectrum (5 times magnified) is recorded with an incident intensity as used in a) and atom density $\rho=6.3\cdot10^{11}~\mathrm{cm}^{-3}$. The spectrum is largely power-broadened and reduced in height due to light scattering which also leads to the plateau around $\delta=0$. This plateau has been included in the simulation as Lorentzian suppression of the signal height. For the red data points the intensity is reduced to $I_\mathrm{in}=0.012\cdot I_\mathrm{sat}$ far below saturation intensity, the atom density is $\rho=5\cdot10^{11}~\mathrm{cm}^{-3}$. Error bars indicate the statistical error of the fitted peak height of four measurements each. For large incident intensity the achieved resolution (blue squares, right axis, red line is a guide to the eye) has been determined from curves as shown in a). In the center where the signal is zero due to light scattering the resolution diverges.}
	\label{fig:Spectra}
\end{figure}
A typical measurement of the reflected light power is shown in Fig.~\ref{fig:Spectra}a). The signal is recorded with a photodiode with $10~\mathrm{kHz}$ low-pass filtering. Its Gaussian shape reflects the density distribution of the atomic cloud in the $z$-direction. The maximum is reached when the center of the atom cloud is reflected at the surface. The Gaussian density distribution of the atoms in the $x$- and $y$-direction, together with probe light that is reflected beside the atom cloud, leads to an averaging of the signal height. For the cloud radii and beam waists used in the experiment the signal in the center of the cloud is expected to be approximately 4 times higher than the averaged signal. During the measurement shown in Fig.~\ref{fig:Spectra}a) approximately $20\%$ of the initially $460.000$ atoms were lost during the reflection, as measured by absorption imaging of the cloud. This loss is explained by the fact that the atom cloud is larger than the gold pad and thus extends beyond the region where the repulsive evanescent wave is plasmonically enhanced. The temperature of the atoms increased due to light scattering during reflection from $1.4~\mathrm{\mu K}$ to $2.0~\mathrm{\mu K}$. The increase was small enough such that the atoms could be trapped in the magnetic trap after the reflection. We note that in this experiment the atoms were reflected by the repulsive action of the probe laser field itself. An additional far-detuned laser for reflecting the atoms as proposed in the theoretical section was not used. For that reason the trajectories of the atoms strongly depend on the detuning and intensity of the probe laser. The measurement shown in Fig. \ref{fig:Spectra}a) was recorded with a detuning of $\delta=12.6\cdot\Gamma$ and an intensity of $I_\mathrm{in}=14.2\cdot I_\mathrm{sat}$, which on the one hand generates a sufficiently high potential barrier for reflecting the atoms, but on the other hand leads to large light scattering. This distorts the corresponding spectrum (black data points) shown in Fig.~\ref{fig:Spectra}b). Small scattering rates are obtained for weak probe light intensities, with which the atoms cannot be reflected from the surface, but are accelerated towards the surface. This is also the case for negative detuning, independently from the intensity. A corresponding recorded spectrum (red data points) for $I_\mathrm{in}=0.012\cdot I_\mathrm{sat}$ is shown in Fig. \ref{fig:Spectra}b). Although this intensity is far below saturation, the fitted theoretical curve corresponds to a linewidth of $13~\mathrm{MHz}$ which is larger than the natural linewidth of $6~\mathrm{MHz}$. The broadening is explained by the non-linearity of $\Delta R(\rho)$ which leads to a reduction of signal height close to resonance.\\ 

The resolution of the detection is determined from curves as shown in Fig.~\ref{fig:Spectra}a). The best signal-to-noise ratio $S_N$ has been achieved for high probe beam intensities. It is calculated as the fitted signal height divided by the noise of the signal. We define the noise as standard deviation, determined from a $4~\mathrm{ms}$ long time track after the atoms are reflected, as indicated in Fig.~\ref{fig:Spectra}a). The atoms that contribute to this signal are confined in the $z$-direction within a thin slice of thickness $l=1/\kappa_3=587~\mathrm{nm}$, while in the radial directions the whole cloud is probed. The number of atoms $N_\mathrm{atom}^\mathrm{det}\approx 9000$ within this slice is calculated from the atomic density. The resolution in atom number shown in Fig.~\ref{fig:Spectra}b) as blue squares is given by $\Delta N=N_\mathrm{atom}^\mathrm{det}/S_N$. Apparently, the resolution diverges for small detuning where the signal is zero due to light scattering. The observed behaviour for negative and positive detuning is different. This is a result of the different atom trajectories. For negative detuning the atoms are accelerated to the surface. This effect is only little influenced by the actual value of the detuning. Furthermore, the signal is reduced for an increasing negative detuning whereas the noise does not change (the used light intensity was the same for all values of detuning). Thus, the resolution is increasing with the detuning. For positive detuning the atoms are reflected at the repulsive potential of the evanescent wave. The minimum distance $z_B$ of the atoms from the surface is reached at the point where the potential height equals the kinetic energy of the atoms. In the experiment this distance varies between $z_B\approx 500~\mathrm{nm}$ at $\delta=120~\mathrm{MHz}$ detuning and $z_B\approx 1.5~\mathrm{\mu m}$ at $\delta=6~\mathrm{MHz}$ detuning. Due to the smaller gap the signal height is enhanced for larger detuning which effectively counteracts the increase of resolution observed for negative detuning. Thus, the resolution for positive detuning is constant, as long as the atoms are reflected from the barrier. With the given parameters the best resolution that we achieved (for positive detuning such that the atoms were reflected) was on the order of $\Delta N=30$ atoms. This resolution is mainly limited by the acoustic noise of the laser power and electronic noise of the photodiode electronics. Moreover, thermal drifts of the setup, in particular of the polarization of the probe ligh field, changed the amplitude of the signal from shot to shot.\\

\begin{figure}[h]
\centerline{\scalebox{1}{\includegraphics{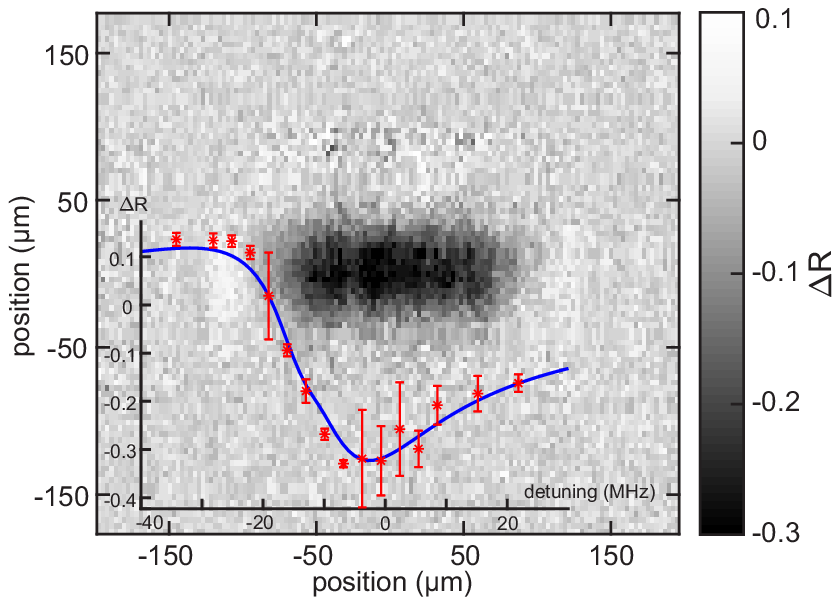}}}
\caption{a) The image of the atom cloud has been taken using the reflected probe beam (atom density $\rho=6.1\cdot10^{12}~\mathrm{cm}^{-3}$, probe laser intensity  $I_\mathrm{in}=0.02\cdot I_\mathrm{sat}$, detuning $\delta=-8.5~\mathrm{MHz}$, and exposure time $400~\mathrm{\mu s}$ ). The experimental resolution is $x_\mathrm{min}=12~\mathrm{\mu m}$. The inset shows a spectrum taken from pixels in the center of the atom cloud.}
\label{fig:image}
\end{figure}

Using the detection scheme we could also take a picture of the slice of atoms at the surface. The reflected light of the probe laser was used to build a $2f$-imaging of the surface with an Ascent Apogee CCD camera. In order to obtain homogeneous illumination of the atom cloud the beam diameter was enlarged to  $\omega_\mathrm{x}=1368\pm10~\mathrm{\mu m} $  and $\omega_\mathrm{y}=1805\pm10~\mathrm{\mu m} $. Thus, the intensity was too low to reflect atoms at the surface. The imaging lens was positioned outside the vacuum chamber which limited the imaging resolution to approximately $12~\mathrm{\mu m}$. The camera was used to record both an image with atoms and a background image without atoms in order to determine the change in reflectivity $\Delta R$. Due to surface roughness including the edges of the gold pads the background of the pictures was obscured by diffraction patterns. We used a fringe removal-algorithm \cite{Ockeloen} to reduce the diffraction patterns on the picture. Fig.~\ref{fig:image} shows a corresponding image of the atom cloud. We note that this is not standard absorption imaging based on the resonant absorption of light, but dispersive imaging that probes the refractive index of the atoms in a submicron thick slice of the cloud. The picture contains the change in reflectivity $\Delta R$ with spatial resolution. This allows us to determine the spectrum of atoms in the center of the cloud without averaging the signal over the transverse directions, see inset of Fig.~\ref{fig:image}. 

\section{Conclusion}
The paper presents a sensitive method for the dispersive detection of ultracold atoms using plasmonically enhanced evanescent waves. As the detection scheme is based on the refractive index of the atoms it is in principle nondestructive. Furthermore, the sensitivity is enhanced by the excitation of surface plasmon polaritons. In typical cold atom experiments, up to several thousands of atoms can be detected with single atom sensitivity and a maximum number of scattered photons of less than one per atom, which makes the method suitable for the quantum nondemolition measurement of matter wave amplitudes. The detection scheme has been experimentally implemented, and atom clouds have been detected with a technically limited resolution of only 30 atoms out of 9000 atoms in total. This resolution is a factor of ten better than what is typically reached by absorption imaging. And there is still room for improvements. Apart from the fact that the light power measurement was not shot-noise-limited, the reached resolution was limited by the fact that the probe light was not only used for detection but also for repelling the atoms from the surface which caused restrictions on the possible range of detuning and light intensity. Using a second far detuned evanescent wave for repelling the atoms would make the detuning and intensity of the probe laser free parameters that can be further optimized for high sensitivity. Another significant improvement could be reached by optimizing the incidence angle in a way that the reflectivity with atoms reaches a minimum. Thus, the shot-noise at the maximum of the signal could be substantially reduced, depending on how small the reflectivity gets. Moreover, a compactification of the setup and active stabilization of the probe light power could help to further reduce acoustic noise in the measured signal. Another possible improvement could be reached by detecting the phase change of the reflected light field due to the presence of the atoms by which the sensitivity can be further enhanced if the thickness of the metal layer is very well controlled \cite{Koenig2008}. Implementing the discussed improvements will push the resolution to the single atom sensitivity. Furthermore, we have extended the detection scheme to probe the refractive index of the atom cloud also locally by imaging a slice of the atom cloud at the surface via the plasmonically enhanced evanescent wave. 

\section{Acknowledgement}
The structured gold surface was fabricated in the plasmonic nanostructures group of Prof. Monika Fleischer, Institute for Applied Physics, University of T\"{u}bingen.
\section{Competing financial interests}
Competing financial interests do not exist.

\end{document}